\documentclass[prd,tightenlines,nofootinbib,showpacs,preprintnumbers,superscriptaddress, twocolumn]{revtex4-1}
\usepackage{amsfonts,amsmath,amssymb,amsthm,bbm,hyperref}
\usepackage{graphicx}
\usepackage{color}
\usepackage{soul}
\usepackage[T1]{fontenc}
\usepackage[utf8]{inputenc}

\newcommand{\be}{\begin{equation}}
\newcommand{\ee}{\end{equation}}
\newcommand{\beq}{\begin{eqnarray}}
\newcommand{\eeq}{\end{eqnarray}}
\newcommand{\nn}{\nonumber}

\begin{document}

\title{Adiabatic Otto-like quantum thermodynamical cycle in the non-quasi-static regime}
\author{Salvador J. Robles-Pérez}
\email{sarobles@math.uc3m.es}
\affiliation{Departamento de Matemáticas, Universidad Carlos III de Madrid. Avda. de la Universidad 30,  28911 Leganés, Spain.}
\author{Salvador Castillo-Rivera}
\email{scastill@math.uc3m.es}
\affiliation{Departamento de Matemáticas, Universidad Carlos III de Madrid. Avda. de la Universidad 30,  28911 Leganés, Spain.}

\date{\today}

\begin{abstract}
We show a finite-time Otto-like quantum thermodynamic cycle that preserves the adiabatic population structure of a time-dependent
harmonic oscillator in the non-quasi-static regime. In the conventional energy representation, finite-rate driving induces non-adiabatic
population redistribution and leaves residual excitations after the
Hamiltonian has returned to its initial value. We show that this
difficulty can be avoided by formulating the dynamics in the
Lewis--Riesenfeld invariant representation, without modifying the
physical Hamiltonian through auxiliary counterdiabatic driving. For a
parametric Mathieu protocol, quantum inertia produces a mismatch
between the spatial width of the working mode and its transient dressed
energy scale. We propose an experimental implementation of this scheme in a trapped-ion Paul trap using stimulated Raman interactions, with independent control
of the laser detuning and beam intersection angle. This provides a finite-time implementation in which the invariant population structure is preserved while the physical trap frequency evolves non-quasi-statically.
Our results establish a clear distinction between adiabatic operation
and quasi-static driving, providing a route toward finite-time quantum
thermal cycles that retain the adiabatic energy structure without
requiring the quasi-static limit.
\end{abstract}

\maketitle



\section{Introduction}

Quantum heat engines  have drawn the attention of the research community as an essential stage in the theoretical formulation of quantum thermodynamics and the study of energy conversion through quantum resources. Some works focus on the heat absorption, emission, and work extraction processes of a quantum Otto cycle \cite{Ishizaki2023}.

The quantum Otto cycle acts as a connection between the macroscopic world of heat engines and the quantum regime of thermal devices of a single component. The model presents the advantage that it is analytically trackable. In fact, an empirical realisation has been accomplished, using a single ion in a harmonic trap \cite{Kosloff2017}.

The quantum Otto engine and refrigeration cycles of a time-dependent harmonic oscillator have been studied by Ref. \cite{Kaur2025}. They researched their optimal performance through a trade-off figure of merit for both adiabatic and nonadiabatic, called sudden-switch, frequency modulations. They reported that the quantum harmonic Otto cycle operated by a sudden-switch protocol cannot perform as a heat engine or refrigerator in the low-temperature boundary. Additionally, they indicated that in the high-temperature limit, the frictional results lead to a complex structure of the phase diagram of the harmonic Otto cycle. 
Rodin \cite{Rodin2024} has proposed a quantum Otto engine based on a three-dimensional harmonic oscillator. One of the modes of the oscillator operates as the working fluid, and the other two emulate baths. The coupling between them is regulated using an external central potential.
Maity and Sen \cite{Maity2025} have developed a protocol for a four-stroke quantum Otto engine that can achieve superior performance when working between two thermal reservoirs. One of them at a positive spin temperature and the other at an effective negative spin temperature. They assumed a procedure that includes a rotating magnetic field in the (x, y)-plane, and a further magnetic field in the z-direction that has distinct strengths. 
Jaramillo et al. \cite{Jaramillo2016} presented a nonadiabatic quantum heat engine running an Otto cycle with a many-particle operating medium, involving an interacting Bose gas contained in a time-dependent harmonic trap. They showed that, through the interaction of nonadiabatic and many-particle quantum effects, the thermal machine may exceed a set of single-particle heat engines with the identical resources, establishing the quantum
supremacy of many-particle thermal machines.
 
Quantum heat engines are set out as thermodynamic cycles with quantum-mechanical functioning media. To achieve elevated engine efficiencies, adiabaticity is needed; a significant issue is to generate a nonvanishing power output at finite cycle times. Shortcuts to adiabaticity (STA) employing counter-diabatic (CD) driving can perform as a means to speed up such infinitely long cycles \cite{Hartmann2020}.

Abah and Lutz \cite{Abah2017} have investigated the performance of a paradigmatic quantum harmonic Otto engine that runs in finite time, as STA techniques are operated to speed up its cycle. The efficiency and power were calculated by accounting for the energetic cost of the shortcut driving. Three different shortcut methods were studied: counterdiabatic driving, local counterdiabatic driving and inverse engineering. As a result, all three led to a concurrent rise of efficiency and power for fast cycles.

The finite-time procedure of a quantum Otto heat engine leads to a trade-off between efficiency and output power. It is driven by the deviation of the system from the adiabatic path. This trade-off limitation can be avoided by operating the shortcut-to-adiabaticity protocol \cite{Shende2024}.

Kim \cite{Kim2021} has shown how time-dependent oscillators execute the STA; first, it employs the setting-up method of an Ermakov-Pinney invariant, and it finds the associated time-dependent oscillator. Afterward, it establishes a class of time-dependent oscillators whose wave functions are known and works out the condition for the STA for adiabatic and nonadiabatic transitions. 

Pedram et al. \cite{Pedram2023} have studied the energetic benefit of accelerating a quantum harmonic oscillator Otto engine through STA (for the expansion and compression stages) and to equilibrium (for the hot isochore), by using CD driving. By contrasting different protocols with and without CD driving. They found that applying both types of shortcut guides improves power and efficiency even after the driving costs are taken into consideration.  

Accomplishing a quick excitation-free quantum control is an essential challenge in current quantum technologies. In numerous cases, STA allow quick adiabatic-like protocols; however, selecting control parameters that meet practical constraints is frequently demanding in complex systems \cite{Xing2026}. 

Karimi and Pekola \cite{Karimi2016} have analysed a quantum Otto refrigerator based on a superconducting qubit connected to two resonators, each of which includes a resistor that operates as a reservoir. They found several regimes: almost adiabatic (low driving frequency), ideal Otto cycle (intermediate frequency), and nonadiabatic coherent regime (high frequency). They observed a substantially improved coefficient of performance compared to an ideal Otto cycle. 

Quantum heat engines demand accurate control over thermal reservoirs and the energies of the quantum operating medium. Notwithstanding that superconducting circuits allow proper engineering of controlled quantum systems, these have not been used to implement a cyclic quantum heat engine. Uusnäkki et al. \cite{Uusnakki2026} have reported on a quantum heat engine with superconducting circuits, employing a quantum-circuit refrigerator as an adjustable heat reservoir and a flux-adjustable transmon qubit as the operating medium. They achieved a few quantum Otto cycles with a tailored reservoir operating to lead to successive cooling and heating, interlaced with flux ramps that controlled the qubit frequency. The outcomes of the work confirmed thermodynamic models of quantum heat engines and advanced control of thermal settings.  

In the view of these considerations, we investigate the performance of a continuous-variable quantum Otto cycle outside the quasi-static regime by analyzing its thermodynamic properties across different quantum representations in a non-quasi-static Mathieu protocol. The outline of the paper is as follows; section II reviews the general quantum properties of a time-dependent harmonic oscillator. Section III establishes the thermodynamic framework of the Otto cycle in the quasi-static limit and discuss its universal operating regimes. Section IV points out the advantages and main inconveniences of using a fast non-quasi-static protocol and the need for shortcuts to adiabaticity. Section V presents a shortcut to adiabaticity based on the invariant representation, and expose the experimental setup using a trapped-ion Paul trap architecture and a stimulated Raman transition. Finally, section VI summarizes the conclusions and discuss the finite-time performance bounds.


\section{Quantum state of a periodic system}\label{QSFloquet}

This work presents an ideal quantum Otto-like cycle based on a harmonic oscillator whose frequency varies periodically between a minimum value, $\omega_m$, and a maximum value, $\omega_M$. We shall not consider friction or dissipative effects, so in the regions where the harmonic oscillator evolves freely, the evolution would be unitary. We shall assume a physical system described by the equation of a harmonic oscillator
\be\label{ho01}
\ddot x +\omega^2(t) x = 0 ,
\ee
with a frequency given by
\be\label{freq01}
\omega^2(t) = a - 2 q \cos 2t ,
\ee
where $a$ and $q$ are two constants. {Mathieu's equation is given by Eq. (\ref{ho01}) and Eq. (\ref{freq01}) and is a classical differential equation. As can be seen,  it is a linear second-order ordinary differential equation with cosine-type periodic forcing of the stiffness coefficient, and it generalizations/ extensions. The extensions involve: geometric nonlinearity, fractional derivative terms, delay terms, quasiperiodic excitation, etc \cite{Landa2012, Kovacic2018}. From simple Josephson circuits and moving to full multimode qubit–cavity systems, Boada et al. \cite{Boada2025} have shown that time-dependent modulation maps the dynamics onto Mathieu-type equations, exposing thresholds for parametric resonances. The transmon qubit is fundamental to quantum computation and displays disordered dynamics under strong parametric drives, which are critical to its control. Enriquez et al. \cite{Enriquez2019} have described an unexplored family of time-dependent single-qubit radiation fields.  These fields are distinguished in terms of the Mathieu functions. They have found that the regions of stability of the Mathieu functions determine the character of the driving fields: for parameters in the stable region, the fields are oscillating and can be periodic under certain conditions. While for parameters in the instability region, the fields are pulse-like. Furthermore, in the stability region, this family reveals solutions for evolution loops in quantum control. Yu et al. \cite{Yu2025} have set out Mathieu control, which uses a non-resonant two-photon drive to provide a particular nonlinear frequency shift. This control supplies a framework for high-fidelity quantum logic and programmable simulation.

The frequency in Eq. \eqref{freq01} is  periodic with period $\pi$ whose value oscillates between the minimum value $\omega_m =\sqrt{a-2q}$, at $t=0$, and the maximum value $\omega_M=\sqrt{a+2q}$, at $t=\pi/2$. The system \eqref{ho01} can be generalized to periodic systems with a different  period by making a change in the variable $t$ and an appropriate rescaling of the constants $a$ and $q$.

The quantum state of a time dependent harmonic oscillator (TDHO) can be given in terms of an orthonormal basis of number states, $\{ |N_I(t)\rangle\}_{N \in \mathbb N}$, whose wave functions are solutions of the Schrödinger equation. They are given (in $\hbar =1$ units) by \cite{Lewis1969, Kanasugui1995, RP2025}
\be\label{WFInv01}
\psi_N^{(I)}(x,t) =  \frac{e^{-i \left( N +\frac{1}{2}\right) \omega_F \tau}}{\sqrt{2^N N! \sigma} }\left( \frac{\omega_F}{\pi} \right)^\frac{1}{4} e^{\left(i\frac{\dot \sigma}{2\sigma} -\frac{\omega_F}{2\sigma^2}\right)  x^2} \text{H}_N(\frac{\sqrt{\omega_F}x}{\sigma}) ,
\ee
where $\omega_F$ is a constant frequency, $\text{H}_N(z)$ is the Hermite polynomial of order $N$, $\dot\tau = \sigma^{-2}$, and $\sigma(t)$ is given by \cite{Strang2005, RP2025}
\be
\sigma(t) = \sqrt{x_1^2(t) + \omega_F^2 x_2^2(t)} ,
\ee
with $x_1(t)$ and $x_2(t)$  two (even and odd) solutions of the Mathieu equation \eqref{ho01} satisfying the initial conditions, $x_1(0) = \dot x_2(0) = 1$ and $\dot x_1(0) = x_2(0) = 0$, and  
\be\label{FF01}
\omega_F\equiv \omega_F(a,q) = \sqrt{\frac{1-x_1^2(\pi;a,q)}{x_2^2(\pi;a,q)}} .
\ee
There are specific values of the parameters $a$ and $q$ for which $\omega_F = \omega(t=0)\equiv \omega_m$. In that case, $|N_0\rangle = |N_I(0)\rangle$, so an initial number state $|N_0\rangle$ evolves as $|N_I(t)\rangle$. Otherwise, the initial state would evolve as a linear combination of number states. It would not introduce any conceptual difference, but it makes the development more obscure. We shall assume throughout this work values of $a$ and $q$ for which, $\omega_F=\omega_m$, so that the initial number state $|N_0\rangle$ evolves as $|N_I(t)\rangle$.

In the instantaneous energy eigenstate basis,
$\{|N_\omega\rangle_{N\in\mathbb N}\}$, the same state can be expressed as a time-dependent superposition,
\be\label{NMOm_01}
|N_0\rangle \rightarrow |N_I(t)\rangle = \sum_M C_M(N;t) |M_\omega(t)\rangle .
\ee
where the coefficient, $C_M(N;t) \equiv \langle M_\omega| N_I\rangle$, can be given in terms of associate Legendre functions \cite{RP2025}. The probability of finding the TDHO in the state $|M_\omega(t)\rangle$ at time $t$ turns out to be \cite{RP2025, Brown1979, Kim1989a},
\be\label{ProbOme_01}
P_M(N;t) = |\langle M_\omega | N_I\rangle|^2  =\frac{M!}{N!} \frac{1}{|\alpha_\omega|} \left| P^\frac{N-M}{2}_\frac{N+M}{2}\left(\frac{1}{|\alpha_\omega|}\right) \right|^2 ,
\ee
provided that $M\pm N$ is an even integer (zero otherwise), and with 
\be\label{alphaOm_01}
\alpha_\omega(t) = \frac{1}{2} \sqrt{\frac{\omega(t)}{\omega_m}} \left( \sigma + \frac{\omega_m}{\sigma \omega(t)}  + \frac{i\dot \sigma}{\omega(t)} \right) \, e^{-i\omega_m\tau}  .
\ee

If the time variation of the frequency is very small, $\frac{|\dot\omega|}{\omega}\ll 1$, we can assume the quasi-static approximation in which $|N_I(t)\rangle \approx |N_\omega(t)\rangle$, and $P_M(N; t) \approx \delta_{MN}$. We can consider that the TDHO stays in the same energy number state along the evolution but with a time dependent  energy of the state that varies according to $\hbar \omega(t)$,
\be\label{energy01}
E(t) = \hbar \omega(t) \left( N +\frac{1}{2}\right) ,
\ee
with $N$ a constant. Overall, the instantaneous energy eigenstates are not solutions of the Schrödinger equation, and the state therefore appears as a redistribution of population among the instantaneous energy levels, so that
\be\label{energy02}
E(t) = \hbar\omega(t) \left(N_\omega(t) + \frac{1}{2}\right),
\ee
with,
\be\label{NOme01}
N_\omega(t) =   \sum_M M P_M(N;t) = \left( |\alpha_\omega|^2 + |\beta_\omega|^2 \right) N + |\beta_\omega|^2 ,
\ee
where, $|\alpha_\omega|^2-|\beta_\omega|^2 = 1$, and $N$ is the quantum label of the initial number state, $|N_0\rangle$.

In terms of the invariant basis, $\{|N_I(t)\rangle\}$, the energy \eqref{energy02} can be written as
\be\label{energy03}
E(t) = \hbar \Omega(t) \left( N +\frac{1}{2}\right) ,
\ee
where $\Omega(t)\geq \omega(t)$ is given by \cite{RP2025}
\be\label{Omega01}
\Omega(t) =  \left( |\alpha_\omega|^2 + |\beta_\omega|^2 \right)   \omega(t).
\ee
with equality in the adiabatic limit, for which $\beta_\omega = 0$.
Here $\Omega(t)$ is an effective energetic frequency associated with the invariant representation, rather than the instantaneous trap frequency $\omega(t)$. Following Eqs.~(10) and (12), the same TDHO state can therefore be described in two equivalent ways: in the instantaneous energy representation, through the time-dependent mean excitation number $N_\omega(t)$ and the instantaneous energy scale $\hbar\omega(t)$, or in the invariant representation, through the constant quantum number $N$ and the effective energetic scale $\hbar\Omega(t)$.


\section{Quantum Otto cycle in the quasi-static approximation}

We propose the model of a quantum thermal machine whose central system, $S$, is represented by a harmonic oscillator with a time dependent frequency $\omega(t)$  that periodically varies from a minimal value $\omega_m$ to a maximum value $\omega_M$. In what follows, we focus on the heat-pump configuration, which is sufficient to establish the finite-time problem and the invariant implementation developed below. In this section, we shall assume that the difference $\omega_M-\omega_m$ is sufficiently small for the harmonic oscillator to undergo a quasi-static evolution along each cycle. Thus, one can consider that the harmonic oscillator stays in the same energy level $|N\rangle$ along the unitary  evolution of the system but with an energy of the energy level that is changing according to Eq. \eqref{energy01}.

With those conditions, the thermodynamic cycle is described as follows (see Fig. \ref{figura3}). The frequency of the harmonic oscillator slowly oscillates between the values $\omega_m$ and $\omega_M$. A conventional heat engine can be experimentally simulated by using the vibrational mode of a single trapped ion as the operating substance. It can be considered to coherently stimulate the ion's vibrational motion as the phonon laser. This aids in acquiring more straightforward results by effectively suppressing thermal fluctuations \cite{Yuan2026}. Overall, lasers perform with a gain medium featuring a large number of pumped systems; lasers have even been assembled at the single-qubit level. Analogous physics to the laser has been found in mechanical oscillators. Such "phonon lasers" have been implemented in a range of systems spanning from atoms to nanomechanics, with the lasing phase displaying a limit cycle of large classical oscillations \cite{Behrle2023}.

The setup contains two electromagnetic beams (lasers), one at frequency $\omega_m$ and the other with frequency $\omega_M$. The central harmonic oscillator is initially in the ground state, where it interacts with the radiation of frequency $\omega_m$, which induces the stimulated excitation of their modes. In the region of absorption  ($1\rightarrow 2$) the state of the harmonic oscillator jumps from $|0\rangle$ to $| 1\rangle$, which is represented by the operator
\be\label{ExcUP01}
\hat P_{\uparrow} = \alpha_{\uparrow} \, |1\rangle \langle 0 | 
\ee
where $\alpha_{\uparrow}$ would quantify the efficiency of the transition process. We shall assume a perfect transition, with $\alpha_{\uparrow} = 1$. If we consider this process instantaneous, the frequency of the harmonic oscillator does not change during the process and the energy change is (in units $\hbar = 1$)
\be\label{energyabsorbed}
\Delta E_{in} = Q_{in} = + \omega_m ,
\ee
where the positive sign in Eq. \eqref{energyabsorbed} means that the harmonic oscillator absorbs the energy from the incident beam of radiation. Eventually, the harmonic oscillator exits the region of absorption and the excitation of modes stops (region $2\rightarrow 3$ in Fig. \ref{figura3}). In the quasi-static approximation, the harmonic oscillator remains in the excited $|1\rangle$ state while the frequency is growing until it reaches the stage $3\rightarrow 4$, where the harmonic oscillator interacts now with the second beam of radiation at frequency $\omega_M$, which stimulates the transition $|1\rangle \rightarrow | 0\rangle$, with an operator
\be\label{ExcDOWN01}
\hat P_{\downarrow} = \alpha_{\downarrow} \, |0\rangle \langle 1 | ,
\ee
for which, as before, we shall assume a perfect efficiency ($\alpha_{\downarrow} = 1$). The process adds one photon  to the beam, with an associated energy increase given by 
\be\label{energyemitted}
\Delta E_{out} = Q_{out} = - \omega_M .
\ee
with, $\omega_M > \omega_m$. The negative sign in Eq. \eqref{energyemitted} means that the harmonic oscillator supplies energy to the radiation beam. Eventually, the harmonic oscillator exits the region of emission and remains in the ground state $|0\rangle$ while the frequency decreases (region $4\rightarrow 1'$) until it enters in the region of absorption, and the cycle starts again.

\begin{figure}

\includegraphics[width=8cm]{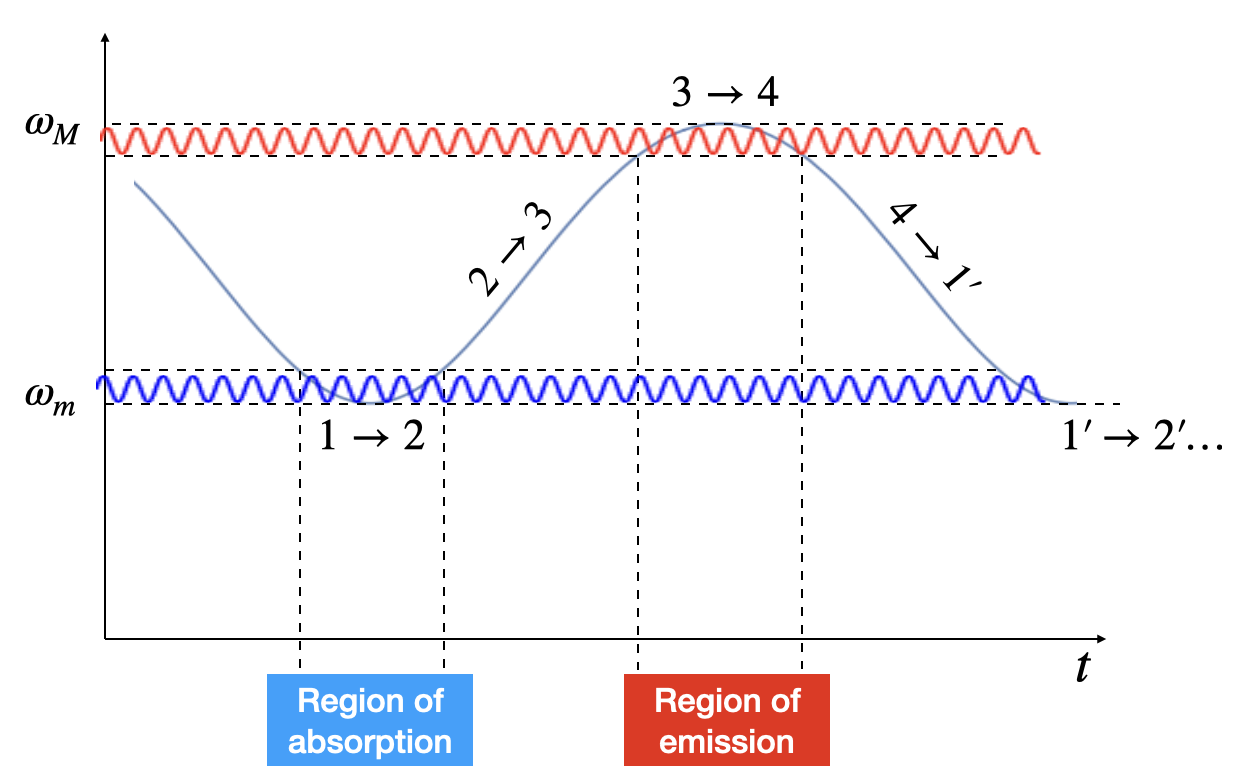}

\caption{The frequency of the central harmonic oscillator during the quantum Otto cycle.}

\label{figura3}

\end{figure}

The quantum Otto cycle operates in close parallel to the classical Otto cycle. Here, the confinement frequency of the central system plays the
role of the piston volume, while the laser fields act as effective reservoirs. The constant-frequency transitions play the role of the isochoric stages, whereas the frequency-modulation stages correspond to adiabatic processes in which the quantum populations remain unchanged. In the quasi-static limit, this cycle maps onto a perfect rectangle in the $(N,\omega)$ plane, which serves as the quantum analogue of the classical $(P,V)$ indicator diagram (see Fig.~\ref{fig::ottocycles}A).

We focus here on the heat-pump operation of the cycle. The work supplied to the working medium over one complete cycle is
\begin{equation}
W_{\rm in} = |Q_{\rm out}|-Q_{\rm in} = \omega_M-\omega_m,
\end{equation} 
where $Q_{\rm in}>0$ denotes the energy absorbed by the working medium
from the low-frequency radiation field, whereas $Q_{\rm out}<0$ denotes
the energy supplied by the working medium to the high-frequency radiation
field. The corresponding coefficient of performance is
\begin{equation}\label{COP_HP_qs}
COP_{\rm HP}^{\rm qs} = \frac{|Q_{\rm out}|}{W_{\rm in}} = \frac{\omega_M}{\omega_M-\omega_m}.
\end{equation}
Although this quasi-static operation reaches the ideal thermodynamic
performance, it requires an infinitely slow modulation. In particular,
the cycle time satisfies
\begin{equation}
\tau_{\rm cycle}\rightarrow\infty,
\end{equation}
so that the finite-time effectiveness, quantified by the trade-off
figure of merit $\chi= COP\cdot \mathcal{P}$, vanishes:
\begin{equation}
\chi_{\rm HP} = COP_{\rm HP}^{\rm qs}
\frac{W_{\rm in}}{\tau_{\rm cycle}}
= \frac{\omega_M}{\tau_{\rm cycle}}
\longrightarrow 0.
\end{equation}
Thus, the quasi-static cycle provides the ideal performance benchmark,
but it cannot deliver finite thermodynamic power. The central problem
is therefore to realize the same adiabatic operation at finite time
without sacrificing its thermodynamic performance. This motivates the
search for \emph{shortcuts to adiabaticity}~\cite{Hartmann2020, Abah2017}.

\begin{figure}

\includegraphics[width=8cm]{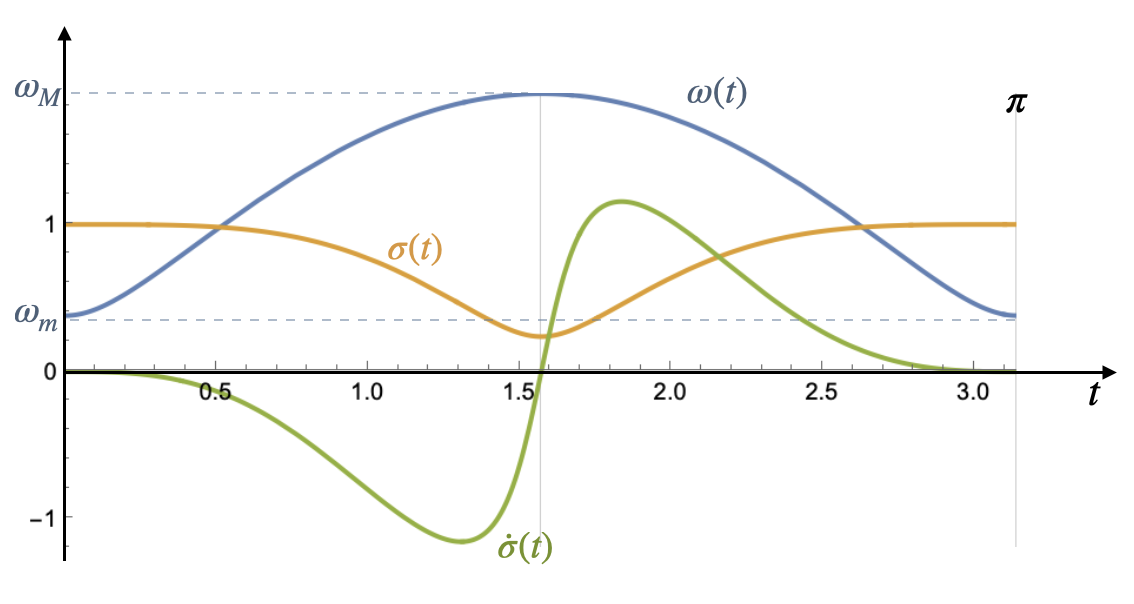}

\caption{The frequency \eqref{ho01}, for $a=1+q$ and $q=0.855$, for which $\omega_F = \omega_m = 0.381$, and the associated $\pi$-periodic values of $\sigma(t)$ and $\dot\sigma(t)$.}

\label{fig_freq}

\end{figure}


\section{Non quasi-static regime and quantum friction}

When the quasi-static approximation is no longer valid, there is a redistribution of the population of the energy eigenstates along the evolution of the TDHO. The analysis has to be done more carefully. We shall do it following the steps given in Fig. \ref{figura3}.

\subsubsection{Step  $1\rightarrow 2$}

The harmonic oscillator is initially in the ground state (of both the invariant and the energy eigenstate basis as they coincide at the initial time, $t_1=0$), 
\be
|\Psi(t_1)\rangle = |0_\omega(t_1)\rangle = |0_I(t_1)\rangle .
\ee
We assume that the interaction between the laser and the harmonic oscillator is in the energy eigenstate basis. Then, the initial state suffers a perfect transition to the $|1_\omega\rangle$ state,
\be
|\Psi(t_1)\rangle  \rightarrow |\Psi(t_2)\rangle = |1_\omega(t_2)\rangle = |1_I(t_2)\rangle ,
\ee
because we have considered the transition occurs instantaneously, so that $t_1 = t_2$. Then, the energy absorbed from the laser and added to the harmonic oscillator is,
\be
\Delta E_{1\rightarrow 2} = \frac{3}{2} \omega(t_2) - \frac{1}{2} \omega(t_1) = + \omega_m
\ee

\subsubsection{Step $2\rightarrow 3$}

If we do not consider dissipative effects, the evolution during this stage is free and unitary. As we have seen in Sec. \ref{QSFloquet}, the state $|1_I(t_2)\rangle$ evolves into a linear combination of energy eigenstates so in the stage $3$ it can be written as,
\be
|\psi(t_3)\rangle = |1_I(t_3)\rangle = \sum_{M\geq 1} C_M(1;t_3) |M_\omega(t_3)\rangle .
\ee
Instead begin given by Eq. \eqref{energy01}, the energy of the oscillator is now given by Eq. \eqref{energy02} or equivalently by Eq. \eqref{energy03}, with $N=1$, so
\be
E(t_3) = \frac{3}{2} \Omega(t_3).
\ee

\subsubsection{Step $3\rightarrow 4$}

The central system now suffers a transition of its energy eigenstates, with $\hat P_{\downarrow}= \alpha_\downarrow |N-1_\omega\rangle \langle N_\omega|$,  for every $N \geq 1$, so that 
\be
|N_\omega\rangle \rightarrow |N-1_\omega\rangle ,
\ee
with perfect efficiency. The quantum state after the transition is then
\be\label{psi4}
|\psi(t_4) \rangle )= \sum_{M\geq 0} C_{M+1}(1;t_3) |M_\omega(t_4)\rangle , 
\ee
where we can consider $t_4 = t_3$ if we assume that the transition occurs instantaneously. The energy of the state in Eq. \eqref{psi4} is given by,
\beq\nn
E(t_4)  &=& \langle \psi(t_4) | \hat H | \psi(t_4)\rangle \\ \nn
&=& \sum_{M\geq 0} |C_{M+1}(1;t_3)|^2 \omega(t_4) \left( M + \frac{1}{2}\right) \\ \nn
&=& \omega(t_4) \left[ \left( \sum_{M\geq 0} P_{M+1}(1;t_3) (M +1) \right) - \frac{1}{2}  \right]  \\ \nn
&=& \omega(t_4) \left( N_\omega(t_3)  - \frac{1}{2} \right)   \\
&=& \frac{3}{2} \Omega(t_3) - \omega(t_4) , 
\eeq
where we have also assumed that\footnote{The probability of transition from the state $|1_I\rangle$ to the ground state  is zero.}, $\sum_{M\geq 0} P_{M+1}(1) =1$, and Eq. \eqref{NOme01}. Therefore, the energy extracted from the harmonic oscillator  is
\be
\Delta E_{3\rightarrow 4} = E(t_4) - E(t_3) = -\omega(t_4) = - \omega_M .
\ee
The total net energy added to the laser in one cycle is
\be
\Delta E = \omega_M - \omega_m .
\ee

\begin{figure*}[t]
\centering
\includegraphics[width=0.98\textwidth]{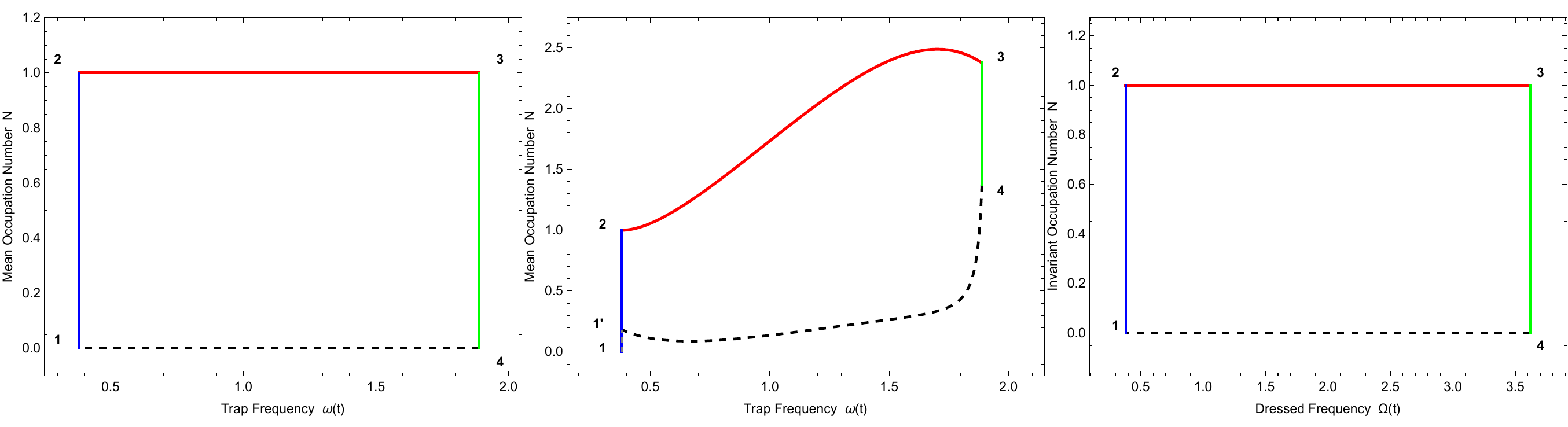}
\caption{Comparative analysis of the quantum Otto cycle across different operation regimes.
(A) The traditional quasi-static regime in the energy representation, yielding the ideal performance limit but vanishing net power in the infinite-cycle-time limit.
(B) The non-quasi-static regime in the energy representation for the first operational cycle. Quantum inertia and non-adiabatic population redistribution deform the cycle path and leave a residual excitation at $1'$, so that the final state does not coincide with the initial ground state $1$, despite the Hamiltonian having returned to its initial value.
(C) The finite-time adiabatic implementation in the Lewis--Riesenfeld invariant representation. The rectangular population structure is preserved because the state-changing interactions are implemented in the invariant basis, while the upper energetic scale is given by the dressed frequency $\Omega(t_3)>\omega_M$.}
\label{fig::ottocycles}
\end{figure*}

\subsubsection{Step $4\rightarrow 1'$}

The state \eqref{psi4} can be written back in the invariant representation as
\be
|\psi(t_4) \rangle = \sum_{M,K} C_{M+1}(1;t_3) \bar C_{K}(M;t_4) |K_I(t_4)\rangle ,
\ee
so the evolution of the TDHO along the step $4$ to $1'$ yields the following state at time $t'_1$
\be\label{psi1p01}
|\psi(t'_1) \rangle = \sum_{M,K} C_{M+1}(1;t_3) \bar C_{K}(M;t_4) |K_I(t'_1)\rangle .
\ee
The energy of this state is obtained by projecting the coherent superposition in Eq.~\eqref{psi1p01} onto the instantaneous energy basis. Since all the parameters return to their initial values at $t'_1$, the invariant and instantaneous energy representations coincide (up to irrelevant phase factors). Therefore,
\begin{equation}
\label{Enet01}
E(t'_1) = \omega_m \left[ \sum_{N\geq0} N \left| \sum_{M\geq0} C_{M+1}(1;t_3)\,\bar C_N(M;t_4) \right|^2 +\frac{1}{2} \right].
\end{equation}
In the quasi-static limit, $|\beta_\omega|\rightarrow 0$, the transition amplitudes become diagonal, $C_{MN}^{(\omega)}(t)\rightarrow\delta_{MN}$ (up to phase factors). Equation~(\ref{Enet01}) then gives $E(t'_1)=\omega_m/2$, as expected for a closed quasi-static cycle. For finite driving rates, however, the transition amplitudes are not diagonal and the state at $t'_1$ generally differs from its initial state, despite the Hamiltonian having returned to its initial value (see Fig.~\ref{fig::ottocycles}B).

\subsubsection{Thermodynamic performance in the non-quasi-static regime}

In the conventional heat-pump operation, the COP is defined as
\be
\mathrm{COP}_{\mathrm{HP}} = \frac{|Q_{\mathrm{out}}|}{W_{\mathrm{net}}},
\ee
where $|Q_{\mathrm{out}}|$ is the heat delivered to the hot reservoir and $W_{\mathrm{net}}$ is the net work supplied over a complete cycle. The quasi-static value is given by Eq.~\eqref{COP_HP_qs}.

Away from the quasi-static regime, the conventional thermodynamic quantities require the explicit evaluation of the energy exchanges along the complete cycle. In particular, the quantum definitions of work and heat depend on the representation used to describe the TDHO~\cite{RP2025}. In the instantaneous energy representation, the work along a frequency-modulation stroke can be evaluated as
\be
W_{i\rightarrow j} = \int_i^j \left(N_\omega(t)+\frac{1}{2}\right)\dot{\omega}(t)\,dt .
\ee
Thus, for the first non-quasi-static cycle,
\be
W_{\rm net}=W_{2\rightarrow3}+W_{4\rightarrow1'},
\ee
where the corresponding integrals require the time-dependent occupation obtained from the transition amplitudes discussed above.

Since the state at $t'_1$ generally differs from the initial state,
the subsequent cycle does not start from the same population
distribution. Repeating the non-quasi-static protocol therefore
requires propagating the population distribution from cycle to
cycle. In some finite-time quantum thermal machines, the repeated
cycle dynamics may converge to a periodic steady state, commonly
referred to as a limit cycle, whose thermodynamic performance can
then be evaluated in the long-time regime~\cite{FeldmannKosloff2012}. However, the
existence and convergence to such a limit cycle are not guaranteed
for arbitrary unitary protocols. Moreover, determining this
long-time state would require iterating the full cycle map, adding
another layer of numerical complexity to the present analysis.

For the purpose of identifying the energetic penalty associated
with finite-time driving, it is therefore more convenient to
introduce the following dimensionless energy-transfer ratio
\be
R_E(\tau) \equiv
\frac{\omega_M-\omega_m}{\Omega(t_3)-\omega_m}.
\ee
Using Eq.~\eqref{Omega01}, this can be written as
\be
R_E(\tau) =
\frac{1}
{1+\left(1+\frac{a}{2q}\right)|\beta_\omega|^2}.
\ee
This quantity equals unity in the quasi-static limit and decreases
as the non-adiabatic excitation generated during the finite-time
stroke increases. Thus, finite-time operation avoids the vanishing
power of the quasi-static limit, but the reduction in $R_E$ signals
the associated energetic penalty. This trade-off motivates the
search for a finite-time implementation that preserves the
adiabatic energy structure without requiring a quasi-static
evolution, as discussed in the following section.



\section{FINITE-TIME ADIABATIC IMPLEMENTATION IN THE
INVARIANT REPRESENTATION}

The results of the previous section show that the loss of
performance at finite driving rates originates from the
non-adiabatic redistribution of the population among the
instantaneous energy eigenstates. Shortcuts to adiabaticity (STA)
provide an established route to suppress such non-adiabatic
excitations and to reproduce adiabatic dynamics in finite time,
typically through suitable modifications of the driving protocol
or by introducing auxiliary control fields~\cite{Guery2019}.
However, these approaches generally require additional control
resources and may introduce energetic and experimental overheads
associated with the auxiliary driving. This motivates the search for
an alternative finite-time implementation that preserves the
adiabatic population structure without introducing an auxiliary
counterdiabatic Hamiltonian.

Here, we take a different approach. Rather than modifying the
Hamiltonian to force the system to follow an adiabatic path, we
implement the state-changing interactions directly in the
Lewis--Riesenfeld invariant representation. In this representation,
the invariant states evolve unitarily without transitions between
the invariant quantum numbers, even when the trap frequency is
driven at a finite rate. Thus, the protocol does not approximate
adiabaticity by means of an auxiliary driving field: the adiabatic
evolution is implemented directly in the representation in which
the invariant quantum number remains constant (see Fig.~\ref{fig::ottocycles}C).

This distinction is important. The time-dependent harmonic-oscillator Hamiltonian underlying
Eq.~(1) remains unchanged; no counterdiabatic term or additional
control field is introduced. What is changed is the representation in which the two
state-changing interactions of the Otto cycle are implemented.
This allows the finite-time dynamics of the TDHO to retain the
adiabatic population structure while the physical trap frequency
still undergoes the non-quasi-static Mathieu modulation.

The implementation of the two isochoric transitions is therefore
changed from the instantaneous energy basis, Eqs. \eqref{ExcUP01}-\eqref{ExcDOWN01},  to the invariant basis.
Specifically, we implement
\be
P_\uparrow = |1_I(t)\rangle\langle 0_I(t)|,
\qquad
P_\downarrow = |0_I(t)\rangle\langle 1_I(t)|.
\ee
At the initial point $t_1$, the state $|0_I(t_1)\rangle$ is mapped
instantaneously onto $|1_I(t_1)\rangle$. The latter then evolves
unitarily as the invariant state $|1_I(t)\rangle$ throughout the
finite-time modulation $2\rightarrow3$, without any redistribution
of its invariant quantum number. At $t_3$, the reverse interaction
maps $|1_I(t_3)\rangle$ onto $|0_I(t_3)\rangle$, which subsequently
evolves as $|0_I(t)\rangle$ during the return stroke.

For the heat-pump configuration considered here, the heat delivered
to the hot reservoir is $|Q_{\mathrm{out}}|=\Omega(t_3)$, while the
net work supplied over the cycle is
\be
W_{\mathrm{net}}=\Omega(t_3)-\omega_m.
\ee
Consequently, the coefficient of performance is
\be
\mathrm{COP}_{\mathrm{HP}}^{\mathrm{inv}} = \frac{\Omega(t_3)} {\Omega(t_3)-\omega_m}.
\ee
The resulting coefficient of performance has the same functional
form as the quasi-static Otto value, with the upper energetic scale
$\omega_M$ replaced by the dressed frequency $\Omega(t_3)$. More importantly, no additional
non-adiabatic excitation term appears in the working medium, because
the invariant quantum number remains unchanged throughout the
modulation strokes. The same adiabatic energy structure is therefore retained at finite
cycle time, with the upper energetic scale replaced by the dressed
frequency $\Omega(t_3)$.

\begin{figure}

\includegraphics[width=8cm]{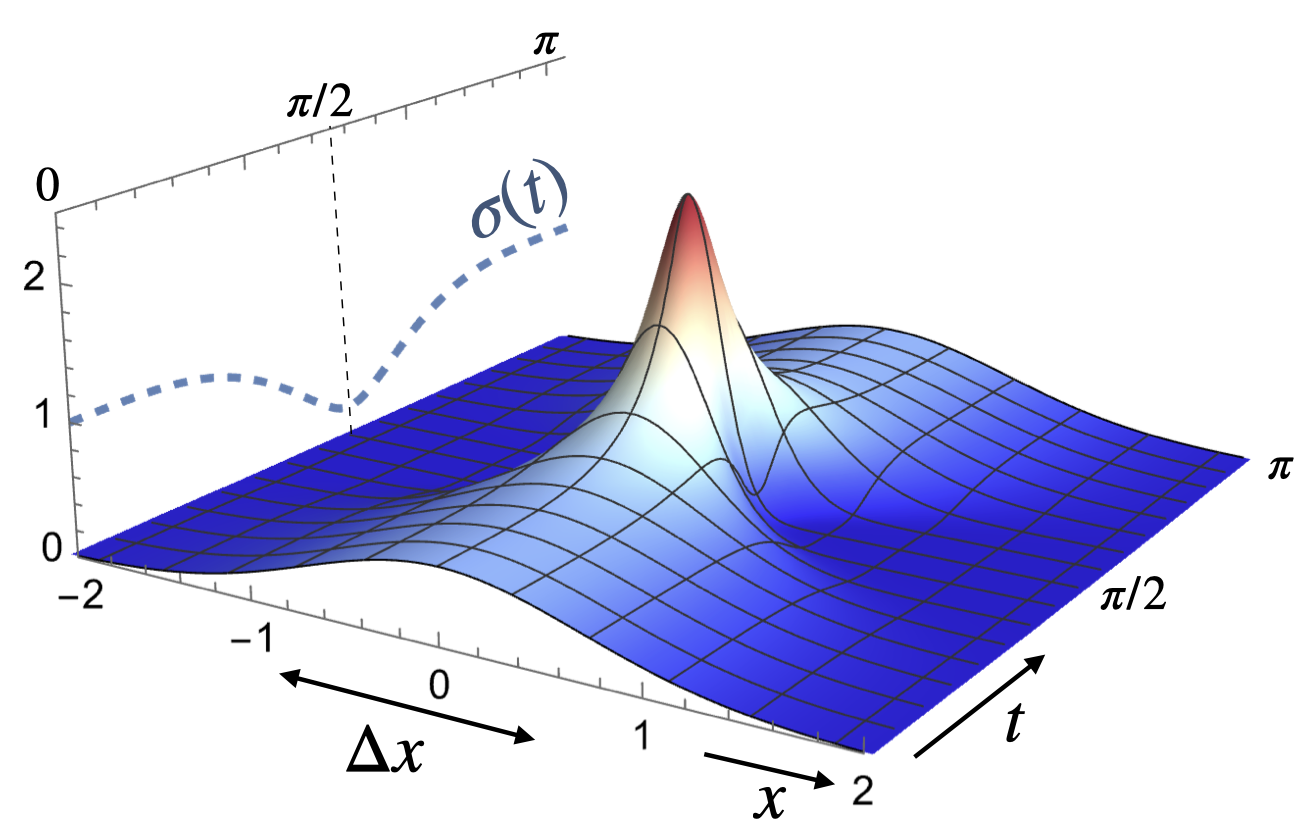}

\caption{Squeezing of the ground state wave packet of the ion, given by $|\psi_0(x,t)|^2$ (see Eq. \eqref{WFInv01}), in a three-dimensional representation. The initially relaxed wave packet with an initial width $\Delta x \sim 1$ (in normalized units) undergoes severe localized compression in the non-quasi-static regime, peaking at the maximum squeezing point $t_3=\pi/2$, according to $\Delta x \sim \sigma(t)$, where the Ermakov parameter $\sigma(t)$ is projected onto the lateral boundary plane ($x=-2$, dashed blue line) showing the direct causal link between the driving potential and the transient geometric narrowing of the quantum state before it unitarily breathes back to its initial configuration.}
\label{fig::mathieusqueezing}

\end{figure}

The resulting finite-time effectiveness is therefore
\be
\chi_{\mathrm{inv}} = \frac{\Omega(t_3)}{\tau_{\mathrm{cycle}}}.
\ee
Unlike the conventional non-quasi-static implementation, no
separate reduction factor analogous to $R_E$ appears: the energetic
scale associated with the invariant state is used directly in the
interaction. Thus, the finite cycle time is not accompanied by the
population redistribution that produced the energetic penalty in
Sec.~IV. In the fast-driving regime, $\Omega(t_3)$ can become much larger
than $\omega_M$, while the cycle time remains finite. The resulting
scaling illustrates that the finite-time implementation does not
require sacrificing the adiabatic energy structure of the cycle.

The invariant implementation, however, introduces a distinct
experimental requirement. At the hot turning point, the energetic
splitting associated with the invariant states is $\Omega(t_3)$,
which in general differs from the instantaneous trap frequency
$\omega_M$ and from the effective spatial frequency governing the
interaction. Let us notice that due to this dynamic excitation, the native boundary frequency of the trap, $\omega_M$, no longer reflects the true energetic splitting between the vibrational states $|0_I(t_3)\rangle$ and $|1_I(t_3)\rangle$. To satisfy strict energy conservation and recover the full energy transfer associated with the dressed energetic scale $\Omega(t_3)$, the laser frequency difference (detuning) must be tuned precisely
to the dressed energetic splitting $\Omega(t_3)$. However, within standard radiation-matter interaction schemes, optical fields couple to the instantaneous eigenstates of the system. In this regard, our model presents an important advantage at the hot turning point $t_3 = \pi/2$: because the time derivative of the Ermakov parameter vanishes identically ($\dot{\sigma}=0$, see Fig. \ref{fig_freq}), the  invariant representation \cite{Lewis1969, Kanasugui1995}
\beq\label{ai01}
\hat a_{I} &=& \sqrt{\frac{\omega_m}{2}} \left( \frac{1}{\sigma} \hat x + \frac{i}{\omega_m} \left( \sigma \hat p_{x} - \dot \sigma \hat x \right) \right) , \\ \label{ai02}
\hat a^\dag_{I} &=& \sqrt{\frac{\omega_m}{2}} \left( \frac{1}{\sigma} \hat x - \frac{i}{\omega_m} \left( \sigma \hat p_{x} - \dot \sigma \hat x \right) \right) ,
\eeq
effectively behaves as the representation of the instantaneously diagonalized basis governed by
\begin{align}
\label{lop01} \hat a_{I} &\approx \sqrt{ \frac{\omega_*}{2}}  \left( \hat x + \frac{i}{\omega_*} \hat p_x \right) , \\ 
\label{lop02} \hat a_{I}^\dagger &\approx \sqrt{ \frac{\omega_*}{2}}  \left( \hat x - \frac{i}{\omega_*} \hat p_x \right) ,
\end{align}
with a frequency given by,
\begin{equation}\label{omega_star_signs}
\omega_* = \frac{\omega_m}{\sigma^2(t_3)} = \frac{\omega_M}{(|\alpha_\omega| - |\beta_\omega|)^2} 
\end{equation}
Nevertheless, a critical mismatch persists since 
\begin{equation}\label{OmegaDressedEq}
\Omega(t_3) = (1 + 2 |\beta_\omega|^2 )\omega_M \neq \omega_*.
\end{equation}
Therefore, a laser tuned to the required energetic scale $\Omega(t_3)$ will completely fail to satisfy the spatial resonance condition demanded by the effective Hamiltonian, meaning the coherent transition cannot be driven.

This energetic--geometric mismatch can be resolved experimentally
by exploiting the independent control provided by a stimulated Raman
transition in a trapped-ion Paul trap~\cite{Leibfried2003}. This experimental architecture offers three distinct advantages for the proposed protocol: (i) the ion is confined along the axial direction by a dynamic radio-frequency (RF) potential that maps onto the time-dependent Mathieu equations considered herein; (ii) the long-lived hyperfine states are virtually immune to spontaneous emission, ensuring quantum coherence over the finite-time durations of the cycle; and (iii) a stimulated Raman configuration enables a complete spatiotemporal decoupling. Following standard frameworks~\cite{Leibfried2003, Bruzewicz2019}, the internal electronic qubit states $\{|g\rangle, |e\rangle\}$ act as a coherent quantum catalyst to drive resonant sideband transitions within the lowest vibrational subspace $N_I=\{0,1\}$. Specifically, the blue sideband drives the coherent transition $|g\rangle |0_I\rangle \rightarrow |e\rangle |1_I\rangle$ implementing $P_\uparrow$ at $t_1=0$, while the red sideband drives the reverse transition $|e\rangle |1_I\rangle \rightarrow |g\rangle |0_I\rangle$ implementing $P_\downarrow$ at the compressed hot stage $t_3=\pi/2$.

In this Raman configuration, the net momentum transfer vector projects exclusively along the axial vibration axis of the trap with a magnitude given by $k_{\text{eff}} = |\vec{k}_1 - \vec{k}_2|$. Assuming $k_1 \approx k_2 \equiv k_L$, the effective spatial wave vector $k_{\text{eff}} = 2 k_L \sin(\theta_H/2)$ is independently adjusted to compensate for the transient spatial squeezing of the wave packet by setting the mutual inclination angle $\theta_H$ between the beams according to\footnote{We assume that the parameters in Eq.~\eqref{ThetaH_Eq} are adjusted to yield a real value for the angle $\theta_H$ to ensure its experimental feasibility.}
\begin{equation}\label{ThetaH_Eq}
\sin\left(\frac{\theta_H}{2}\right) = \frac{\eta_0}{k_L} \sqrt{\frac{\omega_*}{2}},
\end{equation}
where $\eta_0$ denotes the design Lamb-Dicke parameter satisfying the resolved-sideband criterion $\eta_0 \ll 1$~\cite{Leibfried2003}. Thus, the energy exchange at $t_3$ is governed by the dressed
frequency $\Omega(t_3)$, while the spatial interaction remains
maximally efficient through the adjustment of the effective optical
coupling to $\omega^*$.

\begin{figure}

\includegraphics[width=8cm]{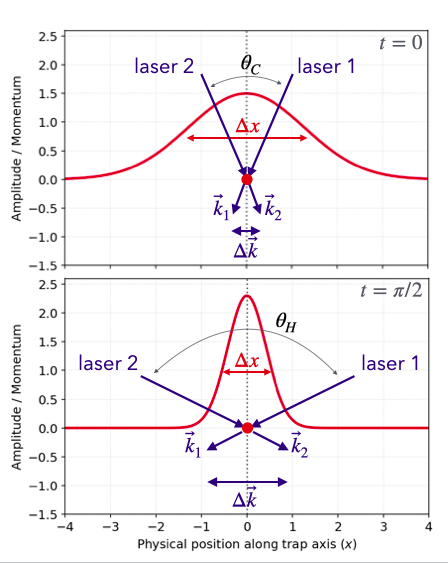}

\caption{Geometric schematic of the two Raman configurations
required to implement the invariant-basis transitions at the two
isochoric turning points. (Top) At the cold point $t=0$, the
unperturbed wave packet is spatially extended and the Raman beams
intersect at the angle $\theta_C$, with the corresponding spectral
detuning $\delta_C$. (Bottom) At the hot point $t=\pi/2$, quantum
squeezing reduces the spatial width of the wave packet. The Raman
configuration is therefore changed to the larger intersection angle
$\theta_H$, while the spectral detuning is simultaneously adjusted
to the dressed energetic splitting $\Omega(t_3)$.}
\label{fig::interaction}

\end{figure}

The implementation described above requires two independently
configured Raman beam pairs, one for each isochoric transition.
This provides the basic physical realization of the protocol without
requiring any additional assumption about the beam-control
architecture. However, for a practical implementation,  the two configurations
can be generated dynamically using a programmable acousto-optic
deflector (AOD). By controlling the RF drive applied to the AOD,
the optical diffraction angle and the corresponding frequency
shift can be modified synchronously, allowing the Raman detuning
and the beam intersection geometry to be switched between the
cold and hot configurations. Phase-locking this RF signal to the parametric Mathieu protocol of the Paul trap allows the quantum Otto engine to operate continuously over multiple consecutive cycles, sustaining finite-time power extraction within the stable coherence window of the hyperfine manifold.

At last, we emphasize how this universal thermodynamic cycle maps onto specific laboratory operations based on our proposed experimental setup. As presented, the phase-locked synchronization of the AOD angles and laser detunings configures the device directly as a high-power quantum heat pump, driving energy into the optical fields (the \emph{thermal} baths). However, due to the fully reversible nature of the invariant shortcut, the exact same architecture can seamlessly operate as a fast quantum refrigerator for ground-state qubit initialization. Experimentally, this refrigeration mode requires inverting the sequence of the blue- and red-detuned Raman sidebands at the turning points, forcing correlated transition operators between arbitrary Fock states $|n_I(t)\rangle \rightarrow |n_I(t) \pm 1\rangle$ across the whole population distribution. In addition, to convert this platform into a work-producing quantum heat engine, the experimental setup would need to be augmented with an extraction channel capable of absorbing the ion's non-quasi-static motional energy without relying on external RF driving. This could be achieved by inductively coupling the ion's axial vibration to an external superconducting LC microcavity or a resonant circuit on the trap chip, effectively extracting the coherent energy absorbed from the lasers as net electrical work.

\section{Conclusions}

We have shown that an Otto-like quantum thermodynamic cycle can retain
its adiabatic energy structure while operating in a finite-time,
non-quasi-static regime. The key is to formulate the dynamics in the
Lewis--Riesenfeld invariant representation, where the relevant
populations remain constant during the driven strokes even though the
physical trap frequency changes rapidly. Importantly, this construction
does not require modifying the physical Hamiltonian by introducing an
auxiliary counterdiabatic term: the finite-time protocol is implemented
through the invariant structure of the original time-dependent
oscillator.

Our analysis of the conventional energy representation clarifies the
physical limitation that arises away from the quasi-static regime.
Finite-rate driving produces non-adiabatic transitions and leaves a
residual excitation after the Hamiltonian has returned to its initial
value, so that the final state $1'$ generally differs from the initial
state $1$. Consequently, subsequent cycles no longer start from the
same population distribution, and their thermodynamic performance
requires following the cycle-to-cycle evolution of the working-medium
state. This highlights the distinction between the first-cycle
energetic penalty and the long-time operation of a finite-time quantum
thermal machine.

The invariant formulation removes this population redistribution from
the driven strokes while retaining the finite duration of the
protocol. The price for finite-time operation is instead encoded in
the dressed frequency $\Omega(t)$, which determines the transient
energetic scale of the working medium. We have further shown how this
separation can be implemented experimentally in a trapped-ion
platform using stimulated Raman interactions, with programmable
control of the relevant optical parameters. The proposed AOD-based
implementation provides a practical route toward synchronizing these
controls with the trap dynamics without introducing an auxiliary
counterdiabatic Hamiltonian.

The resulting framework provides a finite-time realization of an
adiabatic Otto-like cycle and clarifies the apparent tension between
adiabatic operation and non-quasi-static driving. In this sense, the
central result is not to eliminate the energetic consequences of fast
driving, but to separate the adiabatic structure of the state from the
quasi-static evolution of the physical Hamiltonian. This distinction
opens a route toward finite-time quantum thermal machines that preserve
the population structure of adiabatic operation without requiring the
quasi-static limit.

\section*{Acknowledgments}

The work of SJRP was supported by the Grant PID2021-123226NB-I00 (funded by MCIN/AEI/10.13039/501100011033 and by “ERDF A way of making Europe”). Funding for APC: Universidad Carlos III de Madrid  (Agreement CRUE-Madroño 2026)



\bibliographystyle{apsrev4-1}
\bibliography{bibliography}

\end{document}